\documentstyle[12pt,epsbox]{article}

\begin{document}
\rightline{WU-HEP-96-7}
\rightline{BA-TH/96-231}
\vspace{0.3cm}
\begin{Large}
\centerline{{\bf Quantum dephasing by chaos}}
\end{Large}
\vspace{0.5cm}

\centerline{Hiromichi Nakazato, Mikio Namiki, Saverio
Pascazio$^1$}
\centerline{and Yoshiya Yamanaka}
\vspace{0.3cm}

\centerline{Department of Physics, Waseda University, Tokyo 169, Japan}
\vspace{.1cm}
\centerline{$^1$Dipartimento di Fisica, Universit\^^ {a} di Bari and}
\centerline{Istituto Nazionale di Fisica Nucleare, Sezione di Bari}
\centerline{I-70126 Bari, Italy}
\vspace{0.3cm}
\noindent
\centerline{PACS: 03.65.Bz, 24.60.Lz, 03.65.Nk}
\vspace{0.5cm}

\begin{abstract}
We examine whether the chaotic behavior of classical systems with a limited
number of degrees of freedom can produce quantum dephasing,
against the conventional
idea that dephasing takes place only in large systems with a huge number of
constituents and complicated internal interactions. On the basis of this
analysis, we briefly discuss the possibility of defining quantum chaos
and of inventing a ``chaos detector".
\end{abstract}
\vspace{0.3cm}


\newpage

Quantum dephasing is a central issue in quantum measurements.
Many physicists used to think that quantum dephasing
takes place as a result of the interaction with large systems endowed with
a huge number of elementary constituents
and complicated internal interactions. Usually,
some randomness associated with
large systems is considered to play an important role in this context.
See discussions on this matter within the framework of the
many-Hilbert space theory
\cite{mHth}.

It is now known that a class of
classical systems with {\em a few degrees of freedom\/}
behaves chaotically due to nonlinear dynamics \cite{LL}.  This fact
leads us to expect that the interactions between a quantum particle and
a classical system with a limited number of
degrees of freedom, if the latter is in chaotic motion, may give rise to
quantum dephasing on the former. In this note, we
shall investigate the possibility of
such a kind of mechanism (yielding quantum dephasing by chaos arising from
a classical system with a few degrees of freedom),  and shall see from simple
model calculations that this can indeed happen.
We also discuss the issue of quantum chaos
as a natural extension of this kind of argument, and look for a way
to invent a new type of ``chaos" detector.

Let us start by introducing the notion of dephasing from the
mea\-sure\-ment-theoretical point of view. Consider a typical yes-no
experiment of the Stern-Gerlach type as schematically shown in Fig.1.
The whole
measurement process is usually decomposed in two steps: the first being the
spectral decomposition and the second the detection. Each incoming particle
emitted by E and represented by a wave packet $\psi_{0}$ is separated by V
into two branch
waves $\psi_{A}$ and $\psi_{B}$, running in channels A and B,
respectively, corresponding to mu\-tu\-al\-ly-exclusive measurement
propositions ${\cal A}$ and ${\cal B}$.
This is nothing but the spectral-decomposition step in which the
phase correlation be\-tween the two branch waves is kept. This is followed
by the detection step at detector D placed in channel A. Suppose that
$\psi_{A}$ is changed to $\psi_{A}'$ by passing through D. If we have
a coincidence (an anti-coincidence) signal between E and D, we get an
affirmative (negative) answer to ${\cal A}$. Here we have considered
D to be a perfect detector, in the sense that it completely destroys the
phase correlation between the two branch waves $\psi_{A}$ and $\psi_{B}$,
that is, we have perfect dephasing between them.

Suppose that D is simply an instrument, but not necessarily a perfect
detector. In this case, in general, we have a semi-coherent and semi-mixed
case, as will presently be seen.
If the two branch wave packets are guided into the final channel O to
make a superposed state $\psi=\psi_{A}'+\psi_{B}$ (for simplicity, we
have used the same notation for channel O as in channels A and B),
the probability of observing the particle by a perfect detector D$_{0}$,
placed in channel O, reads
\begin{equation}
P=\int_\Omega|\psi|^{2}dx=(\psi_{A}',\psi_{A}')+(\psi_{B},\psi_{B})
+2\mbox{Re}(\psi_{B},\psi_{A}')\ , \label{eqn:P}
\end{equation}
where $\Omega$ is the non-vanishing support of the wave packets.
Usually, we perform a quantum-mechanical observation by sending
many incoming particles (their total number, in an experimental run,
being $N_{p}$) through a steady (and very weak) incident beam
into the target (D in this case), and by accumulating many results
obtained in such a run. If D is a perfect detector, that is, if it
gives perfect dephasing between $\psi_{A}'$ and $\psi_{B}$, the third
term (the interference term) of (\ref{eqn:P}) disappears for the accumulated
distribution. If, on the other hand,
D fully keeps the phase correlation between the branch waves, D$_0$
will read a perfect interference pattern. Otherwise, we obtain an imperfect
measurement, yielding a semi-coherent and semi-mixed case.
This way one can see, by making use of D$_{0}$, whether
D works well or not as a quantum detector.

In a conventional measurement process, each incoming particle will meet,
particle by particle, a different local system.
In other words, the $\ell$\/th particle will interact with the $\ell$\/th
local system and, correspondingly, get a transmission coefficient
$T_{(\ell)}$ when it passes through D. Furthermore, suppose that we can
consider the measurement process as a one-dimensional collision process,
so that we can safely set $\psi_{B}\simeq e^{ikx}$ and $\psi_{A(\ell)}'\simeq
T_{(\ell)}e^{ikx}$, as good approximations. (Notice that $e^{ikx}$ symbolically
represents a wave packet close to a plane wave.)
In this approximation scheme
the accumulated
distribution of results obtained by D$_0$ is given by
\begin{equation}
\overline{P}=1+t+2\sqrt{t(1-\epsilon)}\cos (\arg \overline{T})\ ,
\label{eqn:PT}
\end{equation}
where $t$ and $\epsilon$ are, respectively, the transmission probability
and the {\it decoherence} parameter, defined by
\begin{equation}
t\equiv \overline{|T|^{2}}\quad ,\ \epsilon\equiv 1
-\frac{\,|\overline{T}|^{2}\,}{\overline{|T|^{2}}}\ : \label{eqn:tepsilon}
\end{equation}
with
\begin{equation}
\overline{\cdots} \equiv \frac{1}{N_{p}}\sum_{\ell=1}^{N_{p}}
(\cdots)_{(\ell)}\ .
\end{equation}
Here, of course,
$\epsilon$ is a positive number between 0 and 1, and 
$\epsilon=1$ corresponds to a perfect measurement
(dephasing), $\epsilon=0$ to perfect interference (coherence), and
its intermediate values to imperfect measurements (partially coherent and
partially mixed states).
Thus the value of $\epsilon$ gives us a criterion to judge
how well the instrument D
can work as a measuring apparatus \cite{mHth}.
In particular, the notion of dephasing can be expressed by
\begin{equation}
\mbox{Re}[\overline{\psi'_{A}}\psi_{B}^{*}]=0\ ,\quad \mbox{or}\quad
\overline{T}=0\ ,\quad \mbox{or equivalently}\quad \epsilon=1\ ,
\label{eqn:deph}
\end{equation}
provided that $\overline{|T|^2}\not=0$.

Let us now turn to a different situation from the conventional setup of
the measurement problem, in which the instrument D consists of {\it only one}
particle subject to classical dynamics.
We examine whether this instrument D causes quantum dephasing or not,
and, if yes,
under which circumstances.  For simplicity, we
consider that this classical particle (to be called D-particle)
interacts with an incoming quantum particle via
a potential, its position being
the center of the potential. Furthermore, all recoil effects and
internal structures of the D-particle are neglected.
In such a case, the motion of the $\ell$\/th incoming particle
is described by the potential $V_{(\ell)}\equiv V({\bf r-r}_{(\ell)}(t))$
where ${\bf r}_{(\ell)}$ is the position of the D-particle when the $\ell$\/th
incoming particle meets the D-particle .
{}Our crucial assumption is that the D-particle moves {\em chaotically\/}.
{}Because of the $\ell$-dependence of the chaotic potential center,
we have to add the subscript $_{(\ell)}$ to $P$ and to $\psi_{A}'$
in (\ref{eqn:P}).
As the scatterer consists of a single particle and is not expected
to be large in size, it is not appropriate to treat our present problem
approximately as a one-dimensional collision process.  This means that
the above formula (\ref{eqn:PT}) in terms of $T$ no longer holds,
but we will be able to formulate the {\it decoherence} parameter
in this case along a similar line of thought, as will be shown later.
This new {\it decoherence} parameter preserves its original interpretation:
If dephasing takes place, the {\it decoherence} parameter is equal to unity,
and then the above instrument D is considered to work well
as a quantum detector.

In what follows, for simplicity, we shall suppress the subscript $A$ of the
branch wave. Of course, we shall keep the subscript $_{(\ell)}$ for the branch
wave of the $\ell$\/th incoming particle in channel A.
We can write down the Schr\"{o}dinger equation for the $\ell$\/th
wave packet running in channel A as
\begin{equation}
\left[-\frac{\hbar^{2}}{2m}{\bf \nabla}^{2}+V({\bf r}-{\bf r}_{(\ell)}(t))
\right]
\psi_{(\ell)}({\bf r},t)=i\hbar \frac{\partial}{\partial t}\psi_{(\ell)}
({\bf r},t)\ .
\label{eqn:bseq}
\end{equation}
Note again that ${\bf r}_{(\ell)}(t)$ stands for the center of the
potential at time $t$.
It is easy to see that the translation operator $\exp[\frac{i}{\hbar}
\hat{{\bf p}}\cdot{\bf r}_{(\ell)}(t)]$ reduces the above equation
(\ref{eqn:bseq}) to
\begin{equation}
\left[-\frac{\hbar^{2}}{2m}\nabla^{2}+V({\bf r})+i\hbar {\bf v}_{(\ell)}
\cdot {\bf \nabla} \right]\overline{\psi}_{(\ell)}({\bf r},t)
=i\hbar \frac{\partial}{\partial t}\overline{\psi}_{(\ell)}({\bf r},t)\ ,
\label{eqn:nweq}
\end{equation}
where ${\bf v}_{(\ell)}=\dot{{\bf r}}_{(\ell)}$ and
\begin{equation}
\overline{\psi}_{(\ell)}({\bf r},t)=\exp[\frac{i}{\hbar}\hat{{\bf p}}\cdot
{\bf r}_{(\ell)}(t)]\psi_{(\ell)}({\bf r},t)\ . \label{eqn:nwf}
\end{equation}
In this note, we shall consider only two extreme cases: (i) Adiabatic change, 
and (ii) rapid change.

{\bf Adiabatic change case:} If the center of the potential moves
very slowly during the passage of each incoming wave packet, we can safely
consider the center fixed
in each scattering process.
In this case we can
put ${\bf v}_{(\ell)}=0$, so that
\begin{equation}
\left[-\frac{\hbar^{2}}{2m}{\nabla}^{2}+V({\bf r}) \right]
\overline{\psi}_{(\ell)}({\bf r},t)
=i\hbar \frac{\partial}{\partial t}\overline{\psi}_{(\ell)}({\bf r},t).
\label{eqn:nweq3}
\end{equation}
Here $\overline{\psi}_{(\ell)}$ simply becomes a wave function
for the scattering process
by a fixed potential $V_{0}=V({\bf r})$, whose center is
located at the origin, so that we can
omit the subscript $_{(\ell)}$. (Recall that we are excluding the case in which
the inner motion of the scatterer may give rise to an
additional $\ell$-dependence.)
If we deal with a wave packet very close to a plane wave, we can put
\begin{eqnarray}
\psi_{(\ell)}({\bf r},t)&\simeq& \exp(-\frac{i}{\hbar}E_{k}t)
u_{{\bf k}(\ell)}^{(+)}({\bf r})\ , \label{eqn:stell} \\
\overline{\psi}_{(\ell)}({\bf r},t)&\simeq& \exp(-\frac{i}{\hbar}E_{k}t)
\overline{u}_{{\bf k}}^{(+)}({\bf r}) \ ,
\label{eqn:st}
\end{eqnarray}
during the passage of the wave packet, where $E_{k}=\hbar^{2}k^{2}/2m$ and
\begin{equation}
u_{{\bf k}(\ell)}^{(+)}({\bf r})=\exp(-\frac{i}{\hbar}\hat{{\bf p}}\cdot
{\bf r}_{(\ell)})\overline{u}_{{\bf k}}^{(+)}({\bf r})\ .
\label{eqn:stu}
\end{equation}
Here $u_{{\bf k}(\ell)}^{(+)}({\bf r})$ and $\overline{u}_{{\bf k}}^{(+)}
({\bf r})$ are, respectively, the outgoing solutions of
\begin{eqnarray}
\left[-\frac{\hbar^{2}}{2m}\nabla^{2}+V({\bf r-r}_{(\ell)})\right]
u_{{\bf k}(\ell)}^{(+)}({\bf r})&=&E_{k}u_{{\bf k}(\ell)}^{(+)}
\label{eqn:scuj} \\
\noalign{\noindent and}
\left[-\frac{\hbar^{2}}{2m}\nabla^{2}+V({\bf r})\right]
\overline{u}_{{\bf k}}^{(+)}({\bf r})&=&E_{k}\overline{u}_{{\bf k}}^{(+)}
({\bf r})\ . \label{eqn:scu}
\end{eqnarray}
Note that ${\bf r}_{(\ell)}$ is independent of time and
that $\overline{u}_{{\bf k}}^{(+)}$ has an ${\bf r}_{(\ell)}$-dependent
constant phase in order to match the boundary conditions for
$u_{{\bf k}(\ell)}^{(+)}$ and $\overline{u}_{{\bf k}}^{(+)}$, both of which
are subject to the plane-wave normalization.
Taking into account that
$u_{{\bf k}(\ell)}^{(+)}=W_{(\ell)}u_{{\bf k}}^{(0)}$  and
$T_{(\ell)}=V_{(\ell)}W_{(\ell)}$
($W_{(\ell)}$ and $T_{(\ell)}$ being the
$W$- and $T$-matrices for the potential $V_{(\ell)}$, respectively),
we can write down the scattering amplitude as
\begin{eqnarray}
{}F_{(\ell)}({\bf k}',{\bf k})&=&-\frac{4\pi^{2}m}{\hbar^{2}}
(u_{{\bf k'}}^{(0)},T_{(\ell)}u_{{\bf k}}^{(0)}) \nonumber \\
&=&-\frac{4\pi^{2}m}{\hbar^{2}}\exp[-i{\bf K}\cdot {\bf r}_{(\ell)}]
(u_{{\bf k'}}^{(0)},T_{0}u_{{\bf k}}^{(0)})\ ,
\label{eqn:scamp}
\end{eqnarray}
where ${\bf K}={\bf k'-k}$ stands for the momentum transfer and $T_{0}$
for the $T$-matrix corresponding to the potential $V_{0}$, because
\begin{eqnarray}
V_{(\ell)}&=&V({\bf r-r}_{(\ell)})
=\exp[-\frac{i}{\hbar}\hat{{\bf p}}
\cdot {\bf r}_{(\ell)}]V_{0}\exp[\frac{i}{\hbar}\hat{{\bf p}}
\cdot {\bf r}_{(\ell)}], \label{eqn:VVell} \\
T_{(\ell)}&=&\exp[-\frac{i}{\hbar}\hat{{\bf p}}
\cdot {\bf r}_{(\ell)}]T_{0}\exp[\frac{i}{\hbar}\hat{{\bf p}}
\cdot {\bf r}_{(\ell)}] \label{eqn:TTell}\ .
\end{eqnarray}
These formulas are easily understood on the basis of
the Born approximation
\begin{eqnarray}
{}F_{(\ell)}({\bf k'},{\bf k}) &\simeq& -\frac{m}{2\pi \hbar^{2}}
\int d^{3}{\bf r} e^{-i{\bf K}\cdot {\bf r}}V({\bf r-r}_{(\ell)})
\nonumber \\
&=&-\frac{4\pi^{2}m}{\hbar^{2}}e^{-i{\bf K}\cdot {\bf r}_{(\ell)}}
(u_{{\bf k'}}^{(0)},V_{0}u_{{\bf k}}^{(0)})\ ,
\end{eqnarray}
and its generalization
\begin{equation}
V_{0}\ \longrightarrow\ T_{0}=V_{0}+V_{0}
\frac{1}{E_{k}-\hat{H}+i\epsilon}V_{0}\ ,
\end{equation}
with the total Hamiltonian $\hat{H}= \hat{{\bf p}}^2/2m + V_{0}$.

If we deal with low energy scattering by a short-distance force, we can put
\begin{equation}
{}F_{(\ell)}({\bf k}',{\bf k})=-\exp[-i{\bf K}\cdot {\bf r}_{(\ell)}]kb\ ,
\label{eqn:Fb}
\end{equation}
where $b$ is the scattering length.
%Usually, we can neglect the
%$(\ell)$-dependence of the scattering length.

Along the general line of thought given in \cite{mHth}, we introduce the
{\it decoherence} parameter in a three-dimensional scattering process
\begin{eqnarray}
\epsilon&\equiv& 1-\frac{\,|\int_{\Delta \omega}d\omega\,\overline{F}|^{2}\,}
{\overline{\int_{\Delta \omega}d\omega\,|F|^{2}}} \nonumber
%%\label{eqn:eps3}
\\
&=&1-\left\vert\int_{\Delta \omega}d\omega\,
\overline{\exp[-i{\bf K}\cdot {\bf r}_{(\ell)}]}\,
\bigg/\!\!\int_{\Delta \omega}d\omega\,\right\vert^{2} \label{eqn:eta}
\end{eqnarray}
where $\Delta \omega=(\Delta\theta,\Delta\varphi)$ stands for the solid angle
around $\omega_0=(\theta_0,\varphi_0)$ under which the scatterer sees
the detector.
Clearly this $\epsilon$ serves as a quantitative measure of the degree of
quantum dephasing as in the one-dimensional case.  Notice that this $\epsilon$
depends on $\theta_0$ and $\varphi_0$ in general.

In order to estimate $\overline{\exp[-i{\bf K}\cdot {\bf r}_{(\ell)}]}$,
consider that ${\bf r}_{(\ell)}$ is the resultant point of a
random walk, ${\bf r}_{(1)} \rightarrow {\bf r}_{(2)} \rightarrow
\cdots$, according to the theory of classical chaos \cite{LL}.
Therefore, if the incident beam is steady and very weak,
we can use the Gaussian law with characteristic length $\Delta L$
for the distribution of ${\bf r}_{(\ell)}$, or,
in other words, we can replace the above bar-averaged quantity
$\overline{\exp[-i{\bf K}\cdot {\bf r}_{(\ell)}]}$ with
\begin{equation}
\exp[-N_p\frac{(\Delta L)^{2}}{2}K^{2}]
=\exp[-N_p(\Delta L)^{2}k^{2}(1-\cos \theta)]\ , \label{eqn:GL}
\end{equation}
where $K=2k\sin (\theta/2)$.

The angle-integrals are computed in the following way:
\begin{equation}
\int_{\Delta \omega}d\omega
 =\Delta\varphi\bigl(\cos\theta_0-\cos(\theta_0+\Delta\theta)\bigr)\simeq
\Delta\varphi\Bigl(\Delta\theta\sin\theta_0+
\frac{(\Delta \theta)^{2}}{2}\cos\theta_0\Bigr)\ , \label{eqn:theta}
\end{equation}
\begin{eqnarray}
\lefteqn{\int_{\Delta \omega} e^{-N_p(\Delta L)^{2}k^{2}(1-\cos\theta)}
d\omega}\nonumber\\
&&=\Delta\varphi\,e^{-N_p(\Delta L)^{2}k^{2}}
\int_{\cos(\theta_0+\Delta\theta)}^{\cos\theta_0}
e^{N_p(\Delta L)^{2}k^{2}\xi}d\xi \nonumber \\
\noalign{\vspace{2pt}}
&&=\frac{\Delta\varphi\,e^{-N_p(\Delta L)^{2}k^{2}(1-\cos\theta_0)}}
{N_p(\Delta L)^{2}k^{2}}
\left\{1-e^{-N_p(\Delta L)^{2}k^{2}
(\cos\theta_0-\cos(\theta_0+\Delta\theta))}\right\} \nonumber \\
\noalign{\vspace{2pt}}
&&\simeq\frac{\Delta\varphi\,e^{-N_p(\Delta L)^{2}k^{2}(1-\cos \theta_0)}}
{N_p(\Delta L)^{2}k^{2}}
\left\{1-e^{-N_p(\Delta L)^{2}k^{2}
(\Delta\theta\sin\theta_0+\frac{(\Delta \theta)^{2}}{2}\cos\theta_0)}
\right\}.
\label{eqn:Deltatheta}
\end{eqnarray}
Thus we obtain
\begin{equation}
\epsilon \simeq 1-\left[\frac{e^{-N_p(\Delta L)^{2}k^{2}(1-\cos \theta_0)}
\biggl\{1-e^{-N_p(\Delta L)^{2}k^{2}
(\Delta\theta\sin\theta_0+\frac{(\Delta \theta)^{2}}{2}\cos\theta_0)}\biggr\}}
{N_p (\Delta L)^{2}k^{2}
(\Delta\theta\sin\theta_0+\frac{(\Delta \theta)^{2}}{2}\cos\theta_0)}
 \right]^{2}\ ,
\label{eqn:epslgen}
\end{equation}
from which we conclude that
\begin{eqnarray}
\epsilon &\simeq& 0\ \
\mbox{for}\ \ N_p(\Delta L)^{2}k^{2}
\Bigl(\Delta\theta\sin\theta_0+\frac{(\Delta \theta)^{2}}{2}\cos\theta_0\Bigr)
\ll 1\ ,
\label{eqn:coh} \\
\noalign{\noindent and}
\epsilon &\simeq& 1\ \ \mbox{for}\ \ N_p(\Delta L)^{2}k^{2}
\Bigl(\Delta\theta\sin\theta_0+\frac{(\Delta \theta)^{2}}{2}\cos\theta_0\Bigr)
\gg 1\ .
\label{eqn:deph2}
\end{eqnarray}
We get coherence in the case (\ref{eqn:coh}), and dephasing in
the case (\ref{eqn:deph2}).

Consequently, we arrive at the conclusion that the chaotic motion
of a classical system can generate dephasing, for sufficiently large
$\Delta L$,
even though it has very few degrees of freedom,
or, alternatively, when the classical system has reached a
well-developed (``aged") stage, i.e., $N_p \gg 1$ so that the
replacement (\ref{eqn:GL}) becomes quite reasonable, after the interaction
with many incoming particles.
This suggests the possibility of inventing a new type
of quantum detector, by making use, for instance,
of a ``randomly oscillating mirror".
On the contrary, we observe coherence for very small $\Delta L$,
or when the system is in the developing stage, before ``aging" ($N_p
\simeq 1$), even though in the latter case we have no sound reasoning of the
replacement (\ref{eqn:GL}).

The dependence of the decoherence parameter on $N_p$, the number of particles
in an experimental run, may be an interesting feature in our opinion.
Such a possibility was previously envisaged within the framework of the
many-Hilbert-space approach (see the last paper in \cite{mHth}), but only in
the trivial case of complete dephasing. On the contrary,
Eq. (\ref{eqn:epslgen}) displays a nontrivial $N_p$-dependence.
Such a feature is not known in other theories of measurement,
and in particular is absent in the ``naive" Copenhagen interpretation.
Notice that (\ref{eqn:epslgen}) is a rather general
expression, in that it does not imply any dependence on details of the
interaction (the potential $V$).
Indeed, such a dependence simplifies out in (\ref{eqn:eta}).

{\bf Rapid change case:}
Let us consider the case in which ${\bf r}_{(\ell)}(t)$ in
(\ref{eqn:bseq}) changes very rapidly, during the passage of the $\ell$\/th
incoming wave packet. In this case,
we may first replace the potential term
with
\begin{eqnarray}
\langle V({\bf r-r}_{(\ell)}(t))\rangle&\equiv& \frac{1}{\tau}
\int_{t}^{t+\tau}\!dt\,
V({\bf r-r}_{(\ell)}(t)) \nonumber \\
&=&\int d^{3}{\bf r'} V({\bf r-r'})w_{(\ell)}({\bf r'},t),
\label{eqn:Vw} \\
 w_{(\ell)}({\bf r'},t) &\equiv& \frac{1}{\tau}\int_{t}^{t+\tau}\!dt\,
\delta ({\bf r'-r}_{(\ell)}(t))\ , \label{eqn:wW}
\end{eqnarray}
where $\tau$ is the passage time of the wave packet.
{}Furthermore, let us restrict ourselves to the
situation in which $w_{(\ell)}({\bf r'},t)$
can be replaced
with a statistical distribution ${\cal W}_{(\ell)}({\bf r'-R}_{(\ell)})$
which has width $\sim{\bf \Delta R}_{(\ell)}$ around ${\bf R}_{(\ell)}$
and no time dependence except that
through $\ell$. Under these circumstances, we are allowed to
reduce our scattering problem effectively to that of an average potential
given by
\begin{equation}
{\cal V}_{(\ell)}({\bf r-R}_{(\ell)})=\int V({\bf r-r'}){\cal W}_{(\ell)}
({\bf r'-R}_{(\ell)})d^{3}{\bf r'}\ , \label{eqn:calV}
\end{equation}
leading to the effective Schr\"odinger equation
\begin{equation}
\left[-\frac{\hbar^{2}}{2m}{\bf \nabla}^{2}+{\cal V}_{(\ell)}
({\bf r-R}_{(\ell)})\right]
\psi_{(\ell)}({\bf r},t)=i\hbar \frac{\partial}{\partial t}\psi_{(\ell)}
({\bf r},t)\ .
\label{eqn:bseqrpd}
\end{equation}
It is remarked that in using this equation we are neglecting some sort of
higher-order fluctuation effects on the
Schr\"{o}dinger wave function.

{}For the particular case $V({\bf r})=(2\pi \hbar^{2}/m)b
\delta ({\bf r})$
(i.e.\ Yang's approximation, $b$ being the scattering
length), we further reduce (\ref{eqn:calV}) to
\begin{equation}
{\cal V}_{(\ell)}({\bf r-R}_{(\ell)})=
\frac{2\pi \hbar^{2}}{m} b{\cal W}_{(\ell)}
({\bf r-R}_{(\ell)})
\label{eqn:calVW}
\end{equation}
as a good approximation, because the force range of $V$ is much
shorter than $|{\bf \Delta R}_{(\ell)}|$.

We now find a parallelism between the adiabatic change case and the rapid case
in the above approximation, with correspondence between ${\bf r}_{(\ell)}$ and
$V({\bf r-r}_{(\ell)})$ (see (\ref{eqn:bseq})) in the former and
${\bf R}_{(\ell)}$ and ${\cal V}_{(\ell)} ({\bf r-R}_{(\ell)})$ in the latter.
Note that ${\bf R}_{(\ell)}$ and ${\bf r}_{(\ell)}$ are assumed constant
in both cases. Therefore, we can extend the arguments on the conditions for
quantum dephasing in the adiabatic change case to the present one as well.

It should be remarked that ${\cal W}_{(\ell)}$, in many practical cases,
describes a very dilute and
broad distribution, in which we can regard (\ref{eqn:calVW}) as a constant
potential with strength $(2\pi \hbar^{2}/m)b{\cal W}_{(\ell)}(0)$
(${\cal W}_{(\ell)}(0)\simeq|{\bf \Delta R}_{(\ell)}|^{-3}$) over a
spatial region of a wide spread $|{\bf \Delta R}_{(\ell)}|$.
In this case, we can
easily estimate the scattering phase shift $\chi$ by
\begin{equation}
\chi \simeq -\frac{\lambda b}{|{\bf \Delta R}_{(\ell)}|^{2}}\ .
\label{eqn:phase}
\end{equation}
Here $\lambda$ is the particle wavelength and
$\rho\simeq |{\bf \Delta R}_{(\ell)}|^{-3}$ stands for the scatterer
density. For very large $|{\bf \Delta R}_{(\ell)}|$, this phase shift becomes
very small. This means that we can hardly observe quantum dephasing
in this case. In conclusion, this type of instrument is nothing but
a phase shifter, which can never yield quantum dephasing.

We have so far discarded possible effects caused by the recoil of the
scatterer.  In order to take these recoil effects into account, we just have to
reformulate the scattering amplitude in the above discussion, in an appropriate
way, according to the quantum theory of scattering. In this way,
we can discuss the following two possibilities.

{\bf Quantum chaos:} Consider the case in which both incoming and
target particles are quantum-mechanical. (Note that the target particle
has been treated as a classical particle in the above discussion.)
The formalism described above
still holds if we use the quantum mechanical scattering amplitude
for the collision between incoming
and target particles.
Within this framework, we may be able to reach the
notion of ^^ ^^ quantum chaos", for the {\em target particle state}, via the
observation of ^^ ^^ quantum dephasing" of the scattering amplitude in the
above sense.
On the other hand, N. Saito \cite{ns} suggested that quantum chaos
can arise from possible random phases of the quantum-mechanical
scattering amplitude in the path-integral representation. This idea
may be realized by replacing the $T$-matrix in our formula
(\ref{eqn:scamp}) with a related one in the path-integral
representation, in particular with those in the WKB approximation.
This, as a natural extension of the present approach,
is a promising means to open a doorway into quantum chaos.

{\bf Chaos detector:} If we can detect the above-mentioned recoil effect
as a signal, we have the possibility
of making detectors that contain only a few constituents, for example,
by means of a randomly moving mirror. Detectors of this kind are quite new;
conventional detectors have a huge number of constituents.
Even though we know that the generation and detection of such a signal
poses difficult problems, one such possibilities would be to utilize
the Fourier analysis of the response functions in momentum space.
%\cite{RauchAgarwal}.
If the detector D is characterized by a large value of
the decoherence parameter
$\epsilon$, we can catch the
signal information by observing the Fourier spectrum of
the correlation of the wave functions, defined by $(\psi_B,I(k)\psi'_A)(t)$,
where $I(k)$ stands for
a spectral function yielding the momentum components around $k$.
We expect such a correlation function to depend
strongly on $\epsilon$, in particular when its values are close to unity
(dephasing).

\vspace{0.3cm}

This work was initiated from a preliminary discussion of one of the authors
(M.N.) with N. Saito a few years ago. He thanks Prof.\ N.\ Saito,
Waseda University, for many stimulating discussions, and is also indebted
to Prof.~A.K. Kapoor, University of Hyderabad, for his interest and
hospitality, because the first draft of this paper was
completed in Hyderabad.
The authors are grateful to Prof.\ N.\ Saito for many discussions and
to Prof.\ Y.\ Aizawa, Waseda University, and
Dr.\ S.\ Tasaki, Institute for Fundamental Chemistry (Kyoto), for informing
them of recent studies of chaos . Three of them (H.N., M.N. and Y.Y.) are
partly supported by the Japanese Ministry of Education, Sciences
and Culture. One of them (S.P.) was supported  by the Japanese
Society for the Promotion of Science, under a bilateral exchange program
with Italian Consiglio Nazionale della Ricerche, and by the
Administration Council of the University of Bari.

\newpage

\noindent
{\large{\bf Figure caption}}\\
\begin{description}
\item [Figure 1] Typical yes-no experimental setup of Stern-Gerlach type
\end{description}

%%%\end{document}

\newpage

\pagestyle{empty}

\begin{center}
\epsfile{file=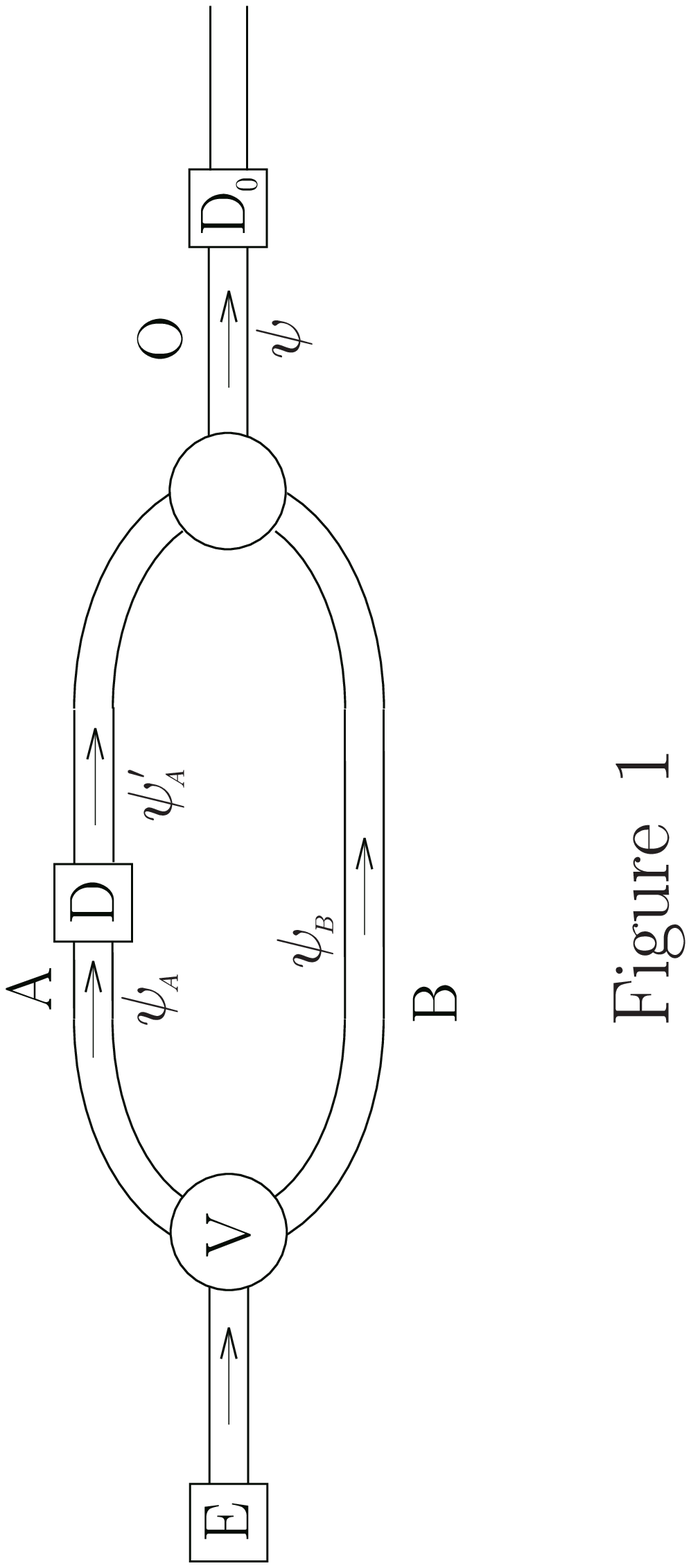}
\end{center}

\end{document}